\documentclass[12pt, english, twoside]{article} 

\usepackage[letterpaper]{geometry}
\usepackage[titletoc,title]{appendix}

\usepackage{fancyhdr} 
\fancypagestyle{plain} 
{
	\fancyhead[LCR]{} 
	\fancyhead[C]{}
	
}

\renewenvironment{abstract}
{\small
	\begin{center}
		\normalsize  \textnormal{ABSTRACT\\} \vspace{-0em}\vspace{0pt}
	\end{center}
	\list{}{%
		\setlength{\leftmargin}{0in}%
		\setlength{\rightmargin}{\leftmargin}%
	}%
	\item\relax}
{\endlist}

\usepackage{sectsty}
\allsectionsfont{\large\raggedright\centering}

\usepackage[]{tocloft}
\addtocontents{toc}{\cftpagenumbersoff{section}} 
\addtocontents{toc}{\cftpagenumbersoff{subsection}} 

\makeatletter
\renewcommand{\@cftmaketoctitle}{}

\renewcommand{\@makefntext}[1]{%
	\setlength{\parindent}{0pt}%
	\begin{list}{}{\setlength{\labelwidth}{6mm}
			\setlength{\leftmargin}{\labelwidth}%
			\setlength{\labelsep}{5pt}
			\setlength{\itemsep}{0pt}%
			\setlength{\parsep}{0pt}%
			\setlength{\topsep}{-3pt}
			\footnotesize}%
		\item[\@textsuperscript{\@thefnmark}\hfil ]#1
	\end{list}%
}

\makeatother

\usepackage{authblk}

\usepackage{graphicx, listings, setspace, hyperref, verbatim}
\usepackage{amsmath, amssymb, mathtools, esint}
\usepackage{tikz, tabularx, makecell, gensymb, adjustbox, subcaption, float}
\usepackage{textcomp}
\usepackage{mathptmx}

\usepackage[T1]{fontenc}
\usepackage[utf8]{inputenc}
\usepackage{mathptmx}

\newcommand\bs{\begin{singlespace}} 			
	\newcommand\es{\end{singlespace}} 		
\newcommand\bq{\begin{quote}\begin{singlespace}\small}	
		\newcommand\eq{\end{singlespace}\end{quote}}
\newcommand\be{\begin{equation}} 			
	\newcommand\ee{\end{equation}}

\let\baraccent=\= 
\renewcommand{\=}[1]{\stackrel{#1}{=}} 

\hypersetup{
	pdftitle={The Toll of the Tolman Effect},
	pdfauthor={Eugene Y. S. Chua and Craig Callender},
	pdfsubject={},
	pdfkeywords={},
	pdfcreator={XeLaTeX},
	pdfproducer={XeLaTeX},
	pdftoolbar=false,	
	pdfmenubar=true,	
	pdffitwindow=false,	
	pdfstartview={FitH},	
	unicode=true,		
	pdfnewwindow=true,	
	colorlinks=true,	
	linkcolor=black,	
	citecolor=black,	
	filecolor=black,	
	urlcolor=blue		
}

\usepackage{filecontents}

\usepackage[main=british]{babel}
\usepackage[babel=true]{csquotes} 

\usepackage[backend=biber, style=ext-authoryear-comp,  maxcitenames=2, dashed=false, url=false, doi=false, eprint=true, giveninits=true, innamebeforetitle=true, maxbibnames=6]{biblatex} 
\DeclareFieldFormat{labeldate}{\mkbibbrackets{#1}} 

\DeclareFieldFormat{postnote}{\mkpageprefix[pagination][\mkcomprange]{#1}} 

\AtEveryBibitem{%
	\clearfield{note}%
	\clearfield{series}%
	\clearfield{number}%
	\clearfield{pagetotal}%
	\clearfield{chapter}%
}

\renewcommand{\intitlepunct}{\addspace\nopunct} 

\DeclareDelimFormat[bib,biblist]{nametitledelim}{\addcolon\space} 

\DeclareFieldFormat{biblabeldate}{\mkbibbrackets{#1}} 

\DeclareFieldFormat{editortype}{\mkbibparens{\textit{#1}}} 
\DeclareDelimFormat{editortypedelim}{\addspace}
\DeclareDelimFormat[bib,biblist]{innametitledelim}{\addcomma\space} 

\DeclareFieldFormat{eprint}{\printtext{available at}\addspace \textless#1\textgreater} 

\DeclareFieldFormat[article,periodical]{volume}{\textbf{#1}} 

\DeclareFieldFormat[inbook, incollection, book]{volume}{\bibcpstring{volume}~#1} 

\DeclareFieldFormat{pages}{\mkpageprefix[pagination][\mkcomprange]{#1}} 

\renewbibmacro{in:}{%
\ifboolexpr{%
test {\ifentrytype{article}}%
or
test {\ifentrytype{inproceedings}}%
}
{}
{\printtext{\bibstring{in}\intitlepunct}}
}

\begin{document}

\title{{\huge The Toll of the Tolman Effect: \\ On the Status of Classical Temperature in General Relativity} \vspace{-1.5ex}}
\author{{\LARGE Eugene Y. S. Chua and Craig Callender  \\ 
\normalsize Accepted at \textit{The British Journal for the Philosophy of Science}. \\ Preprint of 14 July 2025. Please cite published version when available.}}
\date{\vspace{-5.5ex}}

\maketitle 


\noindent\rule{\textwidth}{1pt}

\vspace{5mm}

\begin{abstract}
\noindent The Tolman effect is well-known in relativistic cosmology but rarely discussed outside it. That is surprising because the effect -- that systems extended over a varying gravitational potential exhibit temperature gradients while in thermal equilibrium -- conflicts with ordinary classical thermodynamics. In this paper we try to better understand this effect from a foundational perspective. We make five claims. First, as Tolman knew, it was Einstein who first discovered the effect, and furthermore, Einstein's derivation helps us appreciate how robust it is. Second, the standard interpretation of the effect in terms of `local temperature' leads to the breakdown of much of classical thermodynamics. Third, one can rescue thermodynamics by using Einstein's preferred interpretation in terms of the `wahre Temperatur' -- what we'll call global temperature -- but it too has some costs. Fourth, the effect is perhaps best understood in terms of clocks as opposed to energy loss. Fifth, inspired by a proposal of Einstein's elsewhere, we sketch an interpretation of the effect in terms of a third novel temperature, which we call the `wahre-local temperature'. On this view, temperature -- and thermodynamics -- is defined \textit{only} in relation to local clocks. In sum, we view the fragmentation of temperature in thermodynamics as a natural and expected result of the fragmentation of time in general relativity. \\
\end{abstract}

\tableofcontents

\vspace{5mm}
\noindent\rule{\textwidth}{1pt}

\newpage 

\section{Prehistory} 

In 1873 Francis Guthrie, a mathematician in South Africa, wrote a letter in \textit{Nature} objecting to James Clerk Maxwell’s treatment of gases in a vertical column \parencite{guthrie_kinetic_1873}. \cite{maxwell_theory_1872} had argued that ``gravity produces no effect in making the bottom of the column ... hotter or colder than the top.'' Using intuitive physical arguments, Guthrie held that a system in thermal equilibrium in a gravitational field would have temperature gradients. In their correspondence, Maxwell admits that he changed his mind on this question many times. The issue ``nearly upset'' his ``belief in calculation'' \parencite{maxwell_effect_1990}. Later, Ludwig Boltzmann's teacher and friend Josef Loschmidt argued for the same conclusion Guthrie did, holding that with this temperature gradient the "terroristic nimbus of the Second Law is destroyed." \parencite{loschmidt_uber_1876} Counterarguments by Maxwell and Boltzmann eventually proved decisive. That an equilibrium system's temperature is uniform in a gravitational field was shown by Maxwell, Boltzmann, and Gibbs -- the three giants of statistical mechanics -- and the matter soon closed.  

However, as is well-known in relativistic cosmology but rarely discussed outside it, gravity \textit{does} produce a temperature gradient for systems in equilibrium. That is the conventional interpretation of what is known as the Tolman effect, derived in 1930 by Richard Tolman and Paul Ehrenfest. Inspired by the idea that heat has weight, \textcite{tolman_temperature_1930, tolman_weight_1930} show that for a spatially extended perfect fluid of pure black-body radiation in equilibrium,
\begin{equation}\label{tlocal}
T\sqrt{g_{00}} = const.
\end{equation}
where $T$ is the local rest temperature and $g_{00}$ the time-time component of a static metric. $T$ has metric-dependence; when $g_{00}$ varies spatially, $T$ in equilibrium will also vary spatially, a la Guthrie and Loschmidt but for different reasons.\footnote{ Interestingly, \textcite{loschmidt_uber_1876}'s formula $T = T_0 + az$ (where $z$ is the direction away from Earth) is equivalent to Tolman's in the special case of $a=  -g/c^2.$} Call this temperature $T_L$ for ``local temperature''. In terms of $T_L$, a gas confined to a vertical cylinder on Earth is hotter at the bottom than the top -- and yet still in equilibrium. Because the effect is proportional to $1/c^2$ it may currently be unobservable, but it seems observable in principle. 

What are the ramifications of the Tolman effect? Readers may be surprised to learn that it allows one to break almost every single classical thermodynamic law including one version of the second law, as we'll see; however, it will not permit a \textit{perpetuum mobile}. If we define temperature as $T_L$, most of thermodynamics will have to be subtly rewritten. For example, as readers familiar with thermodynamics will immediately recognize, the effect conflicts with the normal statement of the Zeroth Law, which defines a temperature function on the basis of the uniformity of temperature among the parts of a system in equilibrium. 

With such surprising consequences, why is the Tolman effect not better known? Right now the foundational literature on the effect is small and mostly confined to physics, not philosophical foundations. There exist a handful of conceptual articles,\footnote{See especially the recent \textcite{vidotto2024thermodynamicstime}, which complements our more historical and conceptual discussion by providing an overview and helpful references to the relevant contemporary physics surrounding the Tolman effect.} but these focus mostly on future developments in quantum gravity. Here we want to look more backward than forward. Sticking with classical physics, is the Tolman effect a genuine effect? How should it be understood? And what repercussions does it have for the philosophical foundations of thermodynamics and statistical mechanics? Our paper tackles these questions and makes several novel points and arguments.

First,  in \S2 we emphasize an under-discussed historical point, that \textcite{einstein_doc_1912-1} had already derived the effect well before Tolman -- as Tolman himself knew -- and before general relativity and relativistic cosmology! This is not simply a historical curiosity. Part of the reason the Tolman effect is so little appreciated is that Tolman and Ehrenfest's derivations require many special assumptions. By emphasizing Einstein's simple derivation, we help the reader appreciate that it is an effect like many of the other classic effects of relativity, e.g., gravitational redshift.

Second, the conventional wisdom in physics takes $T_L$ to be the "true" temperature, the counterpart of pre-general relativistic temperature. In \S3 we explain how the Zeroth, First, and Second Laws are all impacted. Maxwell 1868 famously argued that if temperature at equilibrium were not uniform it would allow the creation of a \textit{perpetuum mobile}. We show how the universality of gravity prevents the creation of gravitational Maxwell's demons. 

\begin{figure}[ht]
\centering
\includegraphics[scale = 0.025, alt = {Two vertical blocks. Left block is labeled $T_G$, representing uniform temperature in a gravitational field. Right block is divided into four layers labeled $T_{L4}$ (top) to $T_{L1}$ (bottom), showing temperature variation with height.}]{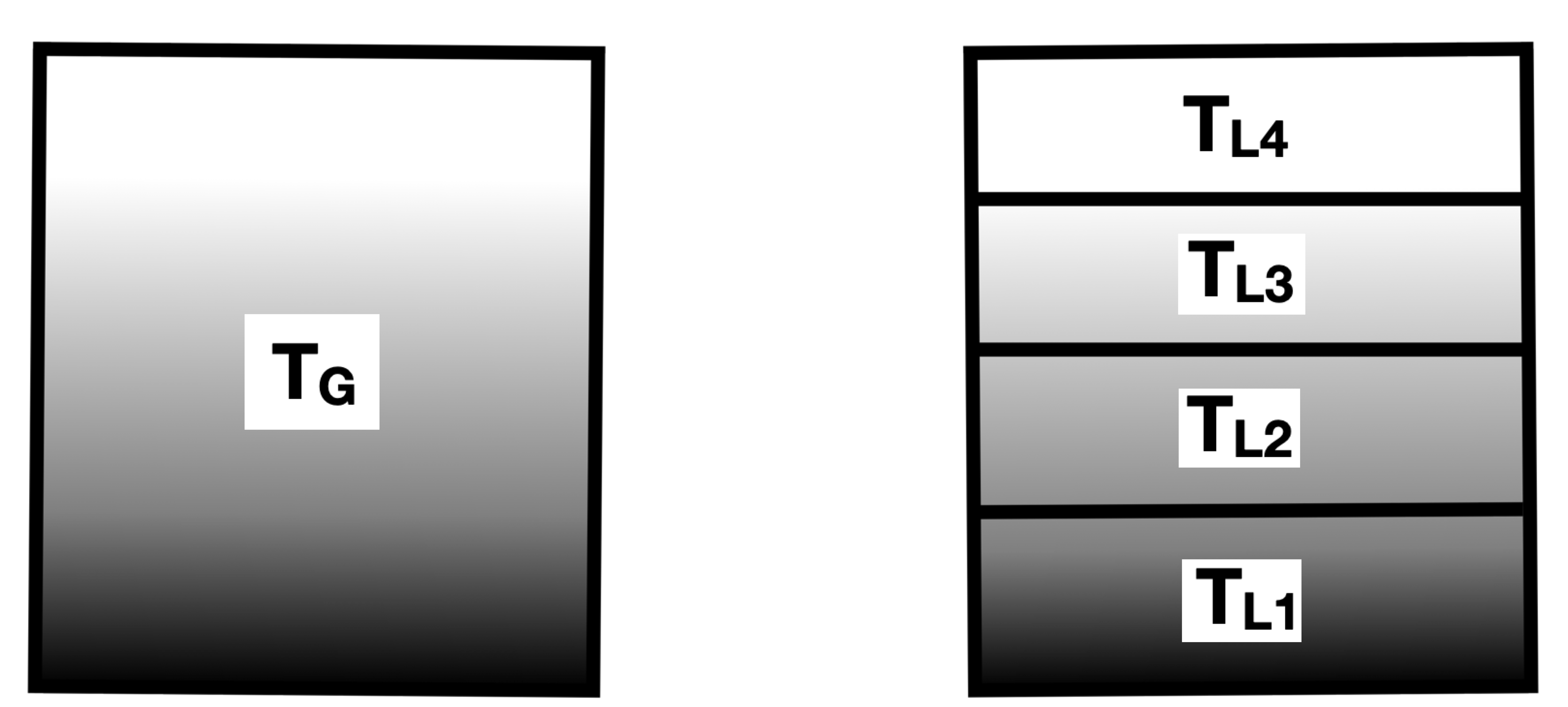}
\caption{$T_L$ vs $T_G$. $T_L$ is constant at each layer (of constant $g_{00}$), but not between layers of varying $g_{00}$. $T_G$ is constant throughout. }
\label{fig:tgvstl}
\end{figure}

Third, is there a way to "save" equilibrium thermodynamics from the Tolman effect? Yes. A natural one is to consider instead a ``global" temperature $T_G$ (see Fig. \ref{fig:tgvstl}), which simply equates temperature with the above constant
\begin{equation}
T_G \equiv const. = T\sqrt{g_{00}}. 
\end{equation}
Using $T_G$ departs from the usual reading of the Tolman effect, but if we use this temperature we can keep the laws of thermodynamics intact. No modifications are necessary. But it too has costs, as we'll see in \S4. The Tolman effect forces a sharper division between observational and theoretical temperatures than we normally encounter. 

Fourth, we try to form a better understanding of the Tolman effect. Motivated by a separate interpretive issue with the gravitational redshift, we identify an "energy interpretation" and a "clock interpretation" of the Tolman effect. Both are given by Einstein. We provide some reasons in favor of a clock interpretation. 

Fifth, this perspective leads to a \textit{third} possible definition of temperature that we've not seen discussed in this context, a definition of temperature Einstein called, elsewhere and much later, the `eigen-temperature': a temperature only defined in local inertial frames at a point. While this recovers all the laws of thermodynamics, this temperature is only defined at a point (plus small region) and amounts to resistance to relativization. 

In sum, the Tolman effect creates a bump in the rug of thermodynamics that no step can entirely remove. While all three temperatures arise from natural considerations based in classical thermodynamics, there is no one natural successor to classical thermodynamics in the general relativistic domain. Already this is a lesson from special relativity (\cite{chua_t_2023}): when one boosts temperature into different inertial frames, the consilience of the classical temperature concept breaks down in special relativity. Here we are essentially boosting temperature into \textit{non}-inertial frames and finding additional fragmentation of the original concept. This highlights how fine-tuned the domain of equilibrium thermodynamics is, and how easily it can fall apart.

\section{The Einstein-Tolman Effect}

The Tolman effect is relatively obscure outside relativistic cosmology. Part of the reason for its inconspicuous status must be due to the way Tolman and Ehrenfest derive the effect in 1930. Although Tolman begins with some intuitive considerations and an Einstein-like argument, he wants to derive the effect exactly in the context of an exact solution to general relativity. The system Tolman uses is a spherically symmetric static perfect fluid. The energy density of this fluid is shown to obey a relativistic Euler equation. Assuming the fluid is pure black body radiation, Tolman is able to connect the energy density to the temperature via the Stefan-Boltzmann law. With this connection he deduces a counterpart of \eqref{tlocal} for this system. With Ehrenfest he then generalized the type of matter used but still worked in the context of relativistic cosmology, using hydrodynamic equations and the Stefan-Boltzmann law. 

While this derivation is insightful and important, it feels precarious due to all the special conditions it invokes. Tolman’s setup has the parts of the system connected by a tube of black-body radiation. Energy conservation is applied to parts but not all of the system. So one wonders whether these ingredients are essential (see \textcite{lima_thermodynamic_2019}). Does it have to be black-body radiation in the tube? Can we disentangle the effect from gravitational redshift? Can we apply energy conservation throughout? Might the effect arise from assumptions that go into the very coarse-grained hydrodynamic equations of relativistic cosmology? All of these questions and more naturally arise, making it hard to appreciate how general or robust the effect is. Its tension with ordinary thermodynamics is immediately evident, yet it is hard to appreciate whether it applies to actual thermodynamic systems familiar from Thermodynamics 101.

For this reason, we speculate, many authors have sought alternative derivations of the effect from more general premises. They have hoped to isolate the essential pieces responsible for the effect. \textcite{balazs_relativistic_1958} and \textcite{balazs_thermodynamic_1965} are a notable and interesting start. They use ordinary thermodynamics plus Einstein’s energy-mass equivalence to derive the effect. \textcite{rovelli_thermal_2011} do similarly, and then speculate on a deeper connection to the thermal time hypothesis. \textcite{santiago_tolman_2019} make a related but distinct argument that relies only on the gravitational redshift for the result and another, also adopted by \textcite{lima_thermodynamic_2019}, that crucially relies on the special relativistic Euler equation. Interestingly, \textcite{santiago_tolman_2019} make a counterfactual claim that ``the Tolman temperature gradient could have been derived some 25 years earlier than it actually was, in 1905 instead of 1930", that is, before the advent of general relativity.\footnote{In fact, in an Appendix they claim that it could have been discovered in 1873. They hold that Nikolay Umov had ``suggestions'' that $E =kmc^2$, and that combined with Euler’s equation would give the Tolman effect. By contrast, we’re assuming $E = mc^2$ comes from Einstein in 1905 and that the principle of equivalence is necessary.} \parencite[5]{santiago_tolman_2019}

Santiago \& Visser are correct. The effect could have been discovered before general relativity because it \textit{actually was}. The history is actual and not counterfactual. As Tolman notes in his 1930 paper, none other than Einstein derives the core effect in 1912, a few years before the discovery of general relativity! And in that derivation \textcite{einstein_doc_1912-1} mentions discussions with Ehrenfest. So we suspect that the core effect was already known to Ehrenfest in 1912 and probably Tolman much before 1930, especially given Ehrenfest's visits to Pasadena and their many collaborations. To them the 1930 papers likely
were important not for the general effect derived but rather for the fact that it could be derived exactly in relativistic cosmology. 

In 1907 Einstein considers the effect of gravitation on clocks and light. He then takes a hiatus from gravity but returns to it in 1911-12. He does not yet have anything like the full theory of general relativity in mind. He hasn’t mastered tensor calculus and tends to lack clarity on many distinctions regarding covariance and invariance \parencite{earman_lost_1978}. He does, however, have an unwavering commitment to a hunch, the principle of equivalence. The principle holds (roughly) that the physics of uniformly accelerating frames and homogeneous static gravitational fields are indistinguishable.\footnote{A must-read on the topic is \textcite{norton_what_1985}; for multiple versions of the principle, see \textcite{okon_does_2011}.} Motivated by this principle, Einstein develops a scalar theory of gravity in which the velocity of light changes as it is modified by the gravitational potential. 

The reader can appreciate the reason why the speed of light changes for him by reflecting on the speed of light in accelerating frames. Using standard radar simultaneity in an accelerating frame, light speed will not be an invariant \parencite{norton_einsteins_2022}. (We return to this when we discuss clocks in \S5.) Invoking the principle of equivalence, Einstein thinks the physical description of the accelerating system should describe what happens in a static gravitational field.  

What are those crucial ingredients? In short, mass-energy equivalence and the principle of equivalence. Already Einstein has $E=mc^2$ in his back pocket from 1905, so he knows that there is an association between energy and inertial mass. Inspired by the equivalence principle he concludes that energy gravitates. This leads him to the ideas that light and heat are affected by gravity. Both have energy, energy is connected to inertial mass via $E=mc^2$, and inertial mass is equivalent to gravitational mass via the principle of equivalence. 

He makes several arguments in 1911, one with a cyclic process he uses in later papers. Suppose we have a light source $S_2$ hanging a height $h$ above a receiver $S_1$ in a uniform
gravitational field. Imagine the source is at the ceiling and the receiver on the floor in a room on the Earth’s surface. We also have a mass M that can move back and forth between $S_2$ and $S_1$, perhaps a weight attached by a pulley. We send some electromagnetic energy $E$ from the ceiling to the floor. The resulting velocity increase due to gravity will increase the electromagnetic pulse’s energy from $E$ to $E(1 + gh/c^2)$. That amount gets absorbed at $S_1$. Now lower the mass $M$ from the ceiling to the floor. That allows work equal to $Mgh$ to be performed. When $M$ is on the floor, give it the extra energy gained from sending the pulse down to the mass; that increases the gravitational mass from $M$ to $M'$. Now lift the mass $M'$ back to the ceiling, $S_1$ — which takes work equal to $M'gh$ — and then transfer the extra energy back to the source, $S_2$. That completes the cycle. 

This cycle threatens to turn into a \textit{perpetuum mobile} of the first kind. One could use the energy gained from dropping the pulse to do work for free — unless conservation of energy
holds, i.e., $M'gh - Mgh = Egh/c^2$. Since we can’t run a perpetuum mobile and conservation does hold, it must be that the difference in gravitational masses, $M'-M$, equals $E/c^2$. Hence gravitational mass changes in accordance with inertial mass, thanks to $E=mc^2$.

With similar reasoning, \textcite{einstein_doc_1912} argues that the energy of a spring is proportional to $c$. Indeed, he claims that the ``same is true of the energy and forces of any systems whatsoever'' \parencite[104]{einstein_doc_1911}. The writing was on the wall for heat energy. 

Finally in a section of \textcite{einstein_doc_1912-1} he turns his attention to heat. Using his now obsolete picture, if the speed of light is $c_0$ at the origin, then the speed at some other location, relative to this origin, is 
\begin{equation}
c = c_0(1 + \phi/c^2)
\end{equation}
where $\phi$ is the gravitational potential at that the location. Einstein’s setup is now two heat reservoirs $W_1$ and $W_2$ at different locations in a static gravitational field. Let $W_2$ be the ceiling and $W_1$ the floor. These two locations have different speeds of light, $c_1$ and $c_2$, 
corresponding to the different locations in a gravitational potential.

Using language he gets from Ehrenfest \parencite[111]{einstein_doc_1912-1}, he calls $T^*$ the `pocket temperature' (or `Taschentemperatur'). He writes that ``the designation ``pocket'' shall be used for any physical apparatus that is regarded as being taken to locations of different gravitational potentials, and the indications of which are always used regardless of the magnitude of c at the location at which they happen to be found'' (111). One is to think of a pocket watch, co-moving with the observer. Einstein had already imagined a pocket light clock and a pocket spring balance. Now we have a pocket thermometer and the associated pocket temperature.

Let’s suppose, Einstein writes, that $W_1$ and $W_2$ have the same pocket temperature. You take a thermometer and measure $W_1$ and then transport the thermometer to $W_2$ and repeat the measurement and get the same result, namely, pocket temperature $T^*$.
With that setup we now perform a simple cyclic process. A body also with pocket temperature $T^*$ is placed against $W_1$ and withdraws heat $Q^*$. The body is then moved to $W_2$ where the heat $Q^*$ is transferred to $W_2$ at temperature $T^*$. Finally the body is moved to reservoir $W_1$, completing the cycle.

What has happened? Einstein is very quick — ``[a]ccording to the results of the previous paper'' — he says it was shown that the heat ``actually withdrawn'' from $W_1$ and supplied to $W_2$ is
\begin{equation}
\begin{aligned}
& Q_1 = Q^*c_1 \\
& Q_2 = Q^*c_2
\end{aligned}
\end{equation}
The previous papers \parencite{einstein_doc_1911, einstein_doc_1912} never mention heat, but in all cases the ``actual'' or ``true'' entity is obtained by multiplying the pocket measure with the local speed of light, which, as we
saw, is connected to the gravitational potential. 

The reasoning is clear. Heat is a form of energy. When we ``drop'' the heat withdrawn from the ``ceiling'' it will gain energy. If we’re not to make a \textit{perpetuum mobile}, energy conservation must hold and the energy gained must be equivalent to the change in gravitational mass of the body moving between ceiling and floor. So the heat difference between the ceiling and floor must be modified by the gravitational potential. 

The Tolman effect is now quickly obtained. Following \textcite[114]{einstein_doc_1912-1}, let's call the unstarred $T$ the `wahre', or true, temperature. This is the temperature that he says is defined by Carnot cycles, that is, the temperature which tracks heat flow in classical thermodynamics. Assuming that the standard thermodynamic relation 
\begin{equation}
\frac{Q_1}{T_1} = \frac{Q_2}{T_2}    
\end{equation}
still holds, we immediately obtain
\begin{equation}
\frac{c_1}{c_2} = \frac{T_1}{T_2}
\end{equation}
so the ratio of wahre temperatures equals the ratio of their speeds of light when both reservoirs have the same pocket temperature $T^*$. Put in the same form in which he presents his pocket clocks and pocket springs, we get 
\begin{equation}\label{wahre}
T = cT^*    
\end{equation}
which is the relationship between the wahre temperature and the pocket temperature.

Consider finally a system at constant $T$ whose parts are thermoconductively connected and yet occupy different locations in a gravitational potential. Then it will suffer a temperature gradient in terms of the pocket temperature $T^*$ that is inversely proportional to the speed of light, which is itself proportional to the local gravitational potential. That is — as clear as day, in 1912 — he discovers the Tolman effect. 

To make that explicit, recall that $c$ for Einstein is equal to $c = c_0(1 + \phi/c^2)$,  so $T = c_0T^*(1 + \phi/c^2)$.\footnote{Note that setting the local speed of light $c_0$ to 1 gives us exactly a common version of the Tolman effect.} In modern notation, the $g_{00}$ components of the metric in a static gravitational field are approximately equal to $(1 + \phi/c^2)$ in the weak field limit. Returning to the two temperatures we introduced in \S1, the pocket temperature is $T_L$ and the wahre temperature is $T_G$, so we arrive at \eqref{tlocal}, precisely the Tolman effect. 

\begin{figure}
\centering
\includegraphics[alt = {Two blocks illustrating the equivalence principle. Left block has an upward arrow labeled `a' (acceleration); right block has a downward arrow labeled `g' (gravity) with grass at the base.}, scale = 0.3]{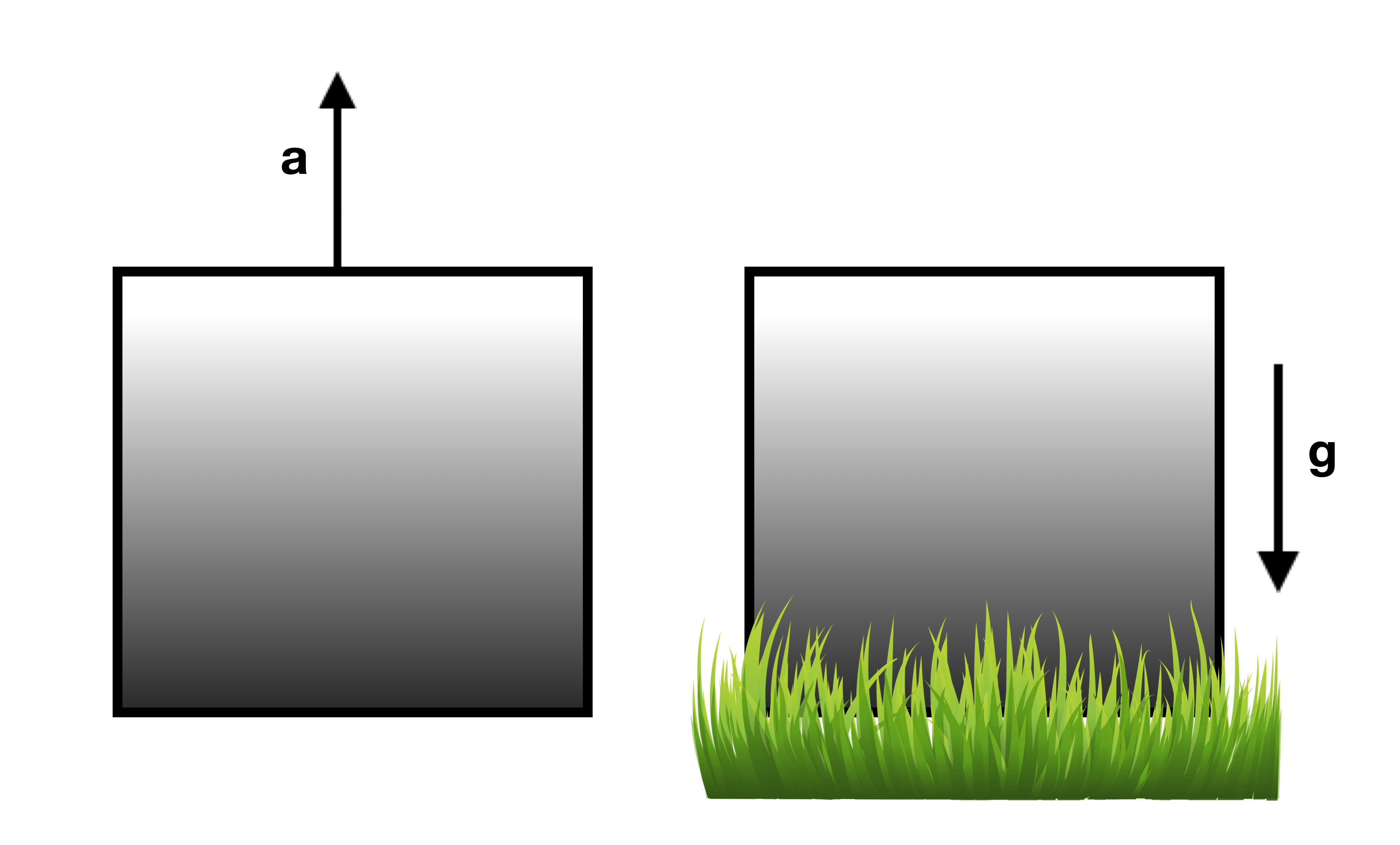}
\caption{The principle of equivalence: a system in a uniformly accelerating frame is roughly indistinguishable from a system in a homogeneous static gravitational field.}
\label{fig:principleofequivalence}
\end{figure}

We can use reasoning in the spirit of Einstein 1911-12 to help independently motivate this conclusion. Einstein’s method in this period is to find the behavior of systems undergoing uniform acceleration and then assume that it also describes a system in a static gravitational field (locally). That suggests a kind of independent check on Einstein's result. Take a relativistic gas in equilibrium and uniformly accelerate it (see Fig. \ref{fig:principleofequivalence}). What will happen to the gas? 

We would expect various gradients to arise due to tidal forces. The parts of the gas in a stronger gravitational field will accelerate faster than the parts in weaker areas, opening up gaps. Still, if the gas obeys the Maxwell-Boltzmann distribution, then we know it will not suffer a temperature gradient. Moving to special relativity, the Maxwell-Boltzmann distribution becomes the Maxwell-J\"uttner distribution \parencite{juttner_maxwellsche_1911}, which also does not create a temperature gradient. That is as expected because the Maxwell-J\"uttner distribution takes into account $E = mc^2$ but not any feature of gravity. When we generalize the Maxwell-J\"uttner distribution to include gravity in static fields or inertial forces, we do obtain the Tolman effect, precisely as expected. Indeed, \textcite{sanchez-rey_thermal_2013} see the Tolman effect in their simulation of a gas obeying relativistic molecular dynamics in Einstein's elevator. Another result in this neighborhood is a study of the Gibbs microcanonical and canonical distributions of a relativistic ideal gas in a uniformly accelerated frame of reference. Once again, the Tolman effect emerges \parencite{louis-martinez_classical_2011}. Using the principle of equivalence, these descriptions of Tolman temperature gradients in accelerating systems must describe what happens (locally) to a gas in equilibrium when sitting in a static gravitational potential. That the effect appears in both kinetic theory and statistical mechanics provides a useful check on our earlier reasoning.

Einstein's scalar theory is a kind of halfway house on the way to general relativity. The reader may wonder how the scalar theory's prediction of temperature gradients compares to that of general relativity. After all, the scalar theory famously predicted the bending of light around the sun, but it only predicted half the deflection angle that the general theory does \parencite{ginoux_albert_2021}. The answer depends on the context and a standard of desired accuracy. Einstein's scalar theory gets the deflection of light around the sun wrong essentially because it ignores the contribution from spatial curvature. It gets the gravitational redshift as measured in the Pound-Rebka experiment (see \S5) right because the experiment happens in a small region with a weak field, allowing us (relative to a standard of measurement) to regard the spatial curvature as irrelevant. In other cases the spatial curvature isn't irrelevant, e.g., measuring gravitational redshift from a distant quasar, just as it isn't when light bends around the sun. Same goes with temperature gradients. In the canonical case, a thermodynamic system sitting in an approximately homogeneous gravitational field, we can introduce the Newtonian potential and approximate the gravitational field as a uniform one, ignoring the spatial curvature. If, however, the object considered was so large that spatial curvature mattered, then the field would be inhomogeneous, and the scalar theory's predictions would depart from Tolman's result in terms of $g_{00}$. Rather than see this departure as a shortcoming of the scalar theory's predictions, we see this as evidence of the \textit{robustness} of the Tolman effect: the effect in scalar gravity proven by Einstein can be seen as an approximation to the more general effect later proven by \textcite{tolman_weight_1930} for the special case of static spacetimes with perfect fluids; this, in turn, can be seen as an approximation to the even more general result proven by \textcite{tolman_temperature_1930}. Recent work suggests it can be generalized even further to non-perfect fluids \parencite{kovtun_temperature_2023}.

The ``thermodynamic'' derivations of \textcite{balazs_relativistic_1958}, \textcite{balazs_thermodynamic_1965}, \textcite{santiago_tolman_2019}, and \textcite{rovelli_thermal_2011} are modern versions of Einstein’s argument -- the last is most faithful to Einstein’s. We agree that the essential ingredients are the energy-mass equivalence and the principle of equivalence linking inertial and gravitational mass. Einstein had both in 1911-12 and they remain core pieces of the general theory today. Readers may prefer other derivations, but we find value in the original. It makes the effect’s reality persuasive and helps us understand -- as we suspect that Tolman and Ehrenfest already knew -- that it does not rely on any especially precarious assumptions
(except those implicit in using the principle of equivalence).\footnote{Since the original derivations, most of the work on the Tolman effect in relativity has been on extending it to broader classes of spacetimes or stress-energy tensors. For example, \textcite{green_dynamic_2014} extend it to stationary, axisymmetric, and asymptotically flat spacetimes, while \textcite{kovtun_temperature_2023} extends it to near-perfect fluids.}

\section{Maxwell’s Vengeance: The Consequences of the Tolman Effect} 

For the moment let’s postpone the discussion of alternative choices of temperature and focus on the local temperature $T_L$. Standard discussions of the Tolman effect almost always interpret it in terms of $T_L$. For example, \textcite[1] {lima_thermodynamic_2019} write:
\begin{quote}
Tolman stressed that the temperature $T$ is directly measurable by \textit{local observers}, and as such, it must be considered \textit{the fundamental quantity} that we mean by temperature at a given point. (emphasis ours)
\end{quote}
Assuming this is correct for now, what does the Tolman effect tell us about thermodynamics?

Einstein and Ehrenfest’s ``pocket temperature'' $T^*$ is the locally observed empirical temperature $T_L$. As David Wallace put it memorably in personal correspondence, it's the temperature measured by any thermometer you can buy at WalMart. Working with this temperature, however, almost all of the traditional laws of thermodynamics are broken in one form or other. Yet, interestingly, a perpetual mobile can't be built to exploit these violations, and by changing temperature concept one can rescue most of the laws.

\textbf{Zeroth Law.} The zeroth law links thermal equilibrium to temperature. Fowler and Guggenheim (1956) coined the term ‘zeroth law’, but its most famous version is given by Planck (1897) as: If $A$ is in thermal equilibrium with $B$, and $B$ is in thermal equilibrium with $C$, then $C$ will be in thermal equilibrium with $A$. The law holds that ‘being in thermal equilibrium’ is an equivalence relation between systems -- which allows for the assignment of an empirical temperature to each system -- and as a consequence that it is also a transitive relation. 

\begin{figure}[ht]
\centering
\includegraphics[alt = {Diagram showing three labeled blocks (A, B, C) at two temperature levels ($T_1$, $T_2$) and two potentials ($\phi_1$, $\phi_2$). Block A sits at $\phi_1$ with $T_1$; block B sits at $\phi_2$ with $T_2$; block C spans both $\phi_1$ and $\phi_2$ with $T_1$ below and $T_2$ above.}, scale = 0.025]{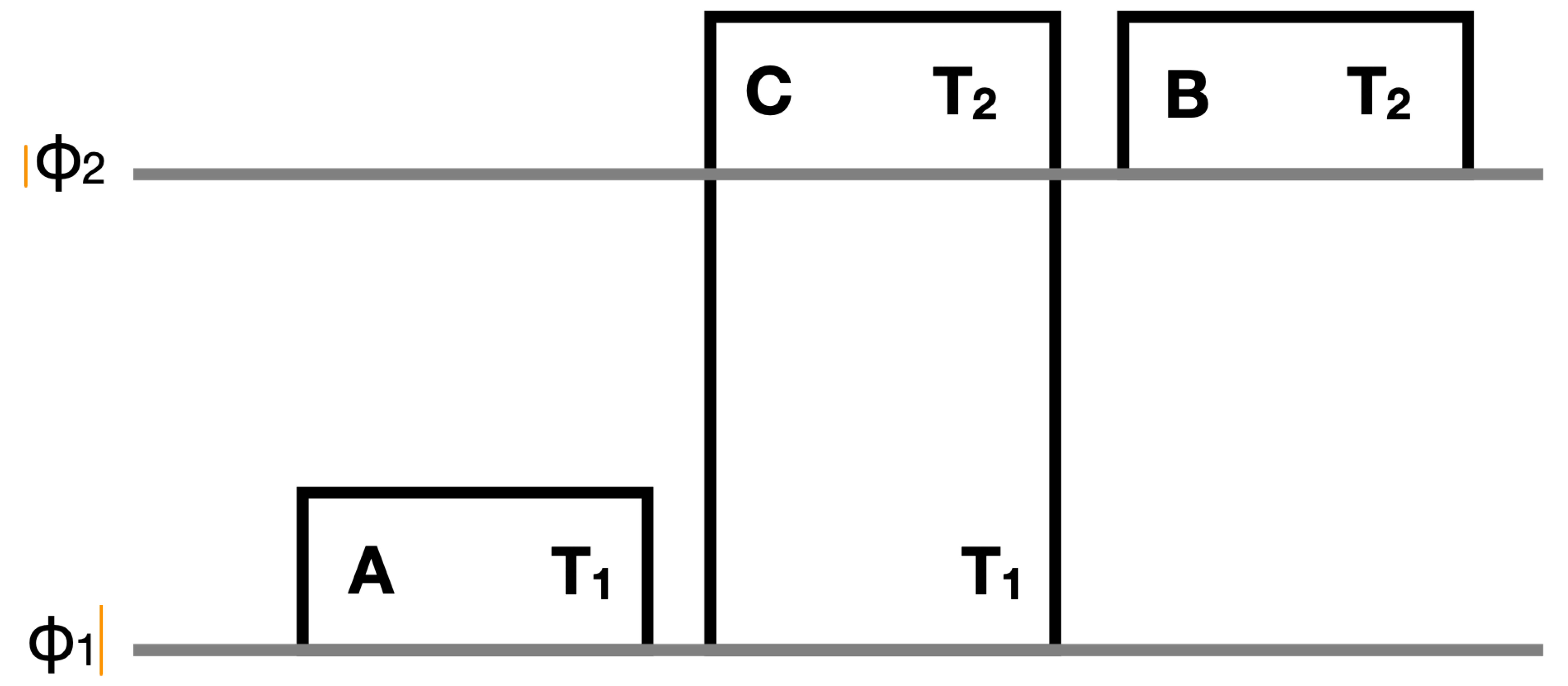}
\caption{Adapted from \textcite{santiago_tolman_2019}.}
\label{fig:santiago}
\end{figure}

The Tolman effect straightforwardly violates the transitivity of same temperature. Here is an example from \textcite{santiago_tolman_2019}. Consider three systems, $A$, $B$, and $C$. $A$ is at height $h_A$ in a
gravitational potential. $B$ is at height $h_B$ in the gravitational potential, where $A \neq B$. And
system $C$ extends from $h_A$ to $h_B$. Due to the Tolman effect, $A$ can be in thermal equilibrium with $C$, $C$ is in equilibrium with $B$, and yet $A$ not in equilibrium with $B$. See Fig. \ref{fig:santiago}.

\textbf{First Law.} The First Law expresses the conservation of energy in thermodynamic terms. The Tolman
effect challenges only the technical statement of the law and not its spirit, as it were.
Mathematically the law states that:
\begin{equation}
dU = dQ + dW
\end{equation}
where $U$ is the internal energy, $Q$ the heat and $W$ work. The $d$ are inexact differentials, which indicates that energy may take the form of heat or work, respectively. This equation must now
include sources of energy from mass-energy equivalence and gravitational energy. But the ``spirit'' of the law still holds, as the law is only "broken" in the sense that more sources of energy now need to be recognized. 

\textbf{Second Law.} The Second Law has many forms. Sometimes advances in physics imply that one form is incorrect in a new regime and yet another one still acceptable. That happens when we consider negative temperatures associated with quantum spin
systems. If we allow negative temperatures, then it turns out that the Kelvin statement of the Second Law requires modification whereas the Clausius statement remains accurate \parencite{lavis_question_2019}. Interestingly, the opposite is the case with the Tolman effect.

The Clausius version of the Second Law states that heat cannot pass from a colder to a warmer body without compensation \parencite{clausius_mechanical_1879}. To run a refrigerator you need to plug it in to an energy source. This is easily violable using the Tolman effect, if `colder' and `warmer' are understood in terms of $T_L$. Put a hotter iron bar, with $T_L^{\text{hot}}$, in contact with a cooler iron bar, with $T_L^{\text{cold}}$, in a gravitational field with varying $g_{00}$ such that the hotter iron bar is at a height with $g_{00}^{\text{hot}}$ and the cooler iron bar is at a height with $g_{00}^{\text{cold}}$. Now, even if $T_L^{\text{hot}} > T_L^{\text{cold}}$, it is still possible that
\begin{equation}
T_L^{\text{cold}} g_{00}^{\text{cold}}  > T_L^{\text{hot}} g_{00}^{\text{hot}}
\end{equation}
if
\begin{equation}
g_{00}^{\text{cold}} > g_{00}^{\text{hot}} \; \; , \; \;  \frac{g_{00}^{\text{cold}}}{g_{00}^{\text{hot}}} > \frac{T_L^{\text{hot}}}{T_L^{\text{cold}}}. 
\end{equation}
That is, it's possible that the gravitational gradient between $g_{00}^{\text{cold}}$ and $g_{00}^{\text{hot}}$ outweighs the temperature gradient between $T_L^{\text{hot}}$ and $T_L^{\text{cold}}$. In such cases, \textit{heat will spontaneously move from the cooler to the hotter bar} until thermal equilibrium is restored and $T_L^{\text{cold}} g_{00}^{\text{cold}}  = T_L^{\text{hot}} g_{00}^{\text{hot}}$, breaking Clausius’ law.

We should quickly reassure the reader that Kelvin’s and other popular expressions of the second law are not violated. There is no gravitational demon. Indeed, it is interesting to see the reason why the Tolman effect cannot be exploited to get free work. Famously, \textcite{maxwell_dynamical_1867} argued that temperature must be the same throughout a system while in equilibrium — on pain of allowing a perpetuum mobile. What is Maxwell's argument and how does the Tolman effect evade it? 

Maxwell’s argument is actually for the universality of uniform temperature at equilibrium \textit{based on} his earlier result in kinetic theory for a vertical column of gas. That result is that the Maxwell distribution of molecular velocities is independent of height in a gravitational field, which implies that temperature is uniform when in equilibrium. Assuming this result, he proves that \textit{any other} system in a vertical column in a uniform gravitational field will also have the same temperature throughout in equilibrium. 

The argument is clever and convincing. Take an ideal gas in a vertical column -- for which we know temperature is uniform -- and place it on a thermally conducting hot plate. Take any other system of any kind of substance and place it also on the hot plate. Then if there are any temperature differences at any heights in that system we can make a \textit{perpetuum mobile}, he claims. It would be easy. Simply connect a thermally conducting rod or other connector between the two systems at the top. Then, in his words, ``if the temperatures of the tops of the two columns were different, we might drive an engine with this difference of temperature, and the refuse heat would pass down the colder column, through the conducting plate, and up the warmer column; and this would go on till all the heat was converted into work, contrary to the second law of thermodynamics.'' Normally, that is precisely right. 

\textcite{loschmidt_uber_1876} considered this very argument but did not accept the premise that kinetic theory already established that temperature was uniform in an ideal gas. That demonstration, he said, assumed that particles would travel in straight lines, but under a gravitational force they should instead travel parabolas. However, Loschmidt also knew that we could modify the original argument. With \textcite{santiago_tolman_2019}, we can run a version with only one column. Just attach the rod from one height to another. The temperature gradient can be used to drive an engine. If we have refuse heat we send it back in until all the heat is converted into work, contrary to the second law. That seems pretty compelling. Loschmidt, however, didn't try to modus tollens the modus ponens and accepted that ``inexhaustible resource of convertible heat at all times".  

How does Tolman’s effect avoid this consequence? The answer is that nothing shields gravity. On the Tolman effect, temperature is affected point by point
by the local gravitational potential. The temperature gradient will therefore apply also to the rod, wire, or any device used to connect the different locations of the system. At each level surface of constant height in any system on Earth, the temperature will be the same in equilibrium. To rig a gravitational demon, one needs something to transport heat from one layer to another. But in equilibrium the temperature along the transport device will itself change temperature (i.e., get hotter as the bar is lower), so the transport device will deliver to the lower level a substance at precisely the same temperature as obtains at that height.

Gravity’s universality stops the gravitational demon from forming. Interestingly, one can mimic this scenario with electromagnetism and two differently charged gases. Then the \textit{non}-universality of electromagnetism substantially alters the diagnosis \parencite{santiago_tolman_2019}.

Returning to the main thread, although the Tolman effect will not allow a \textit{perpetuum mobile}, it will wreak havoc with the laws of equilibrium thermodynamics as we normally conceive of them.
The zeroth, first, and some versions of the second laws are all violated. Many subsidiary laws and principles are also straightforwardly affected, e.g., Fick’s law, how temperature governs heat flow in Carnot cycles, how there's no heat flow under a constant-temperature adiabatic process, and so on. Interestingly, the minus-first law — that systems will spontaneously move to equilibrium — is not affected (on this, more below).

\section{Wahre is the True Temperature?}

The conventional wisdom assumes $T_L$ is the real or best counterpart of the classical temperature of thermodynamics. However, since the effect's discovery it has been known that there is a choice. \textcite[912]{tolman_weight_1930} operated entirely in terms of ``the proper temperature, as measured by a local observer". Yet Ehrenfest may have had an influence on Tolman in their sequel paper, when \textcite[1798]{tolman_temperature_1930} write that: 
\begin{quote}
...it might be emphasized that although the proper temperature itself, $T_0$, varies from point to point in a gravitational system which has come to equilibrium, nevertheless the constancy of the combined quantity $T_0 \sqrt{g_{00}}$ provides many of the advantages of the older principle of constant temperature throughout as necessary for equilibrium. Indeed it would be possible to label $T_0 \sqrt{g_{00}}$ as \textit{the} temperature of a system, except for the undesirability of multiplying the different things that are signified by that word. In this connection it is also interesting to recall that Einstein himself was led in his early speculations on the nature of gravitation to distinguish between a quantity, called \textit{``wahre Temperatur,''} which would be constant throughout a system in thermal equilibrium and a second quantity, called at the suggestion of Ehrenfest \textit{``Taschentemperatur,''} which would vary with gravitational potential.
\end{quote} 
Indeed, Einstein credits Ehrenfest in 1912 with providing him the "pocket" ("Taschen") terminology that he applies to the local temperature. 

Put in modern language, the idea is to simply define a ``global" temperature $T_G$, as
\begin{equation}\label{tglobal}
T_G \equiv const. = T_L \sqrt{g_{00}}. 
\end{equation}
While $T_L$ is defined in terms of the usual internal energy $U$, the global temperature is defined in terms of the total energy $U_G$ which includes gravitational potential energy. It is the former that exhibits a gradient due to the Tolman effect, while the latter remains constant even under the Tolman effect. This distinction between the wahre Temperatur and the Taschentemperatur maps nicely onto our distinction between $T_L$ and $T_G$. 

The global temperature is possible to define because we've assumed a background static gravitational field. The geometry for such a field possesses a global time-like Killing field. With respect to this field a global energy (and associated conservation law) can be defined in terms of the Lie derivative. That allows us to define a total energy $U_G$ and hence also $T_G$. (Recall that the Tolman effect was discovered via application of the principle of equivalence, which is itself limited to homogeneous static gravitational fields.) Outside this context it's unclear whether there can be an appropriately defined global temperature just as there is a question of whether global energy can be appropriately defined in general relativistic spacetimes without the appropriate symmetries.\footnote{This is a question we put aside for this paper. For discussion, see e.g. \cite{hoefer_energy_2000}, \cite{maudlin_status_2020}, \cite{duerr_against_2021}, \cite{ramirez_causation_nodate}.}

We will not argue that one needs to choose between temperature concepts. So long as one is clear about which concept is being used, there should be no problem. Still, even if one adopts this "conventionalist" perspective, some conventions are better or more natural than others. By noting the various strengths and limitations of each concept, we also learn about how the ordinary temperature concept breaks down in different ways in a gravitational context. There is a sense in which one cannot "have it all" with any temperature concept. 

The main virtue of $T_L$ is that it is a locally observable quantity. When taking a good thermometer and placing it against a system, the empirical temperature that you see on the thermometer is $T_L$. It is primarily for this reason that the conventional reading of the Tolman effect is in terms of the local temperature and hence of temperature gradients. The main problem, as we saw, is that the choice of  $T_L$ wreaks havoc on thermodynamics as ordinarily understood. 

This revisionary result is not intolerable. After all, we have a compelling explanation of why this happens and why we never noticed. Ordinary thermodynamics arose from developing steam engines and the like in weak near-earth gravitational fields of roughly the same potential, contexts wherein the Tolman effect couldn't possibly manifest itself observationally. It was perfectly natural to ignore the gravitational potential in the thermodynamic internal energy. As physics developed, we've since learned that thermodynamics relies on many special assumptions and that it is not strictly true in all regimes, e.g., at the scale of Brownian motion. Same here.  

$T_G$ has some virtues too. To us, two stand out. The first, of course, is that $T_G$ is defined so as to be constant in equilibrium. Hence all the issues that arose in the previous section disappear. The second virtue, we think, is deeper and more interesting. Some physicists and philosophers of physics have come to understand the laws of thermodynamics to sit atop a more basic tendency in nature, one often confused with the Second Law. The tendency is that systems out of thermal equilibrium spontaneously relax to unique states of thermal equilibrium. This tendency is sometimes called the Minus First Law \parencite{uhlenbeck_lectures_1963, brown_origins_2001, myrvold_explaining_2020}. We like this picture of thermodynamics because it cleanly separates "thermostatics" -- the time-independent aspects of classical equilibrium thermodynamics -- from the time-dependent dynamical underpinning of the theory. 

One adopting this view of thermodynamics might be tempted to say that whatever temperature makes sense of the Minus First Law is really the best temperature counterpart. After all, it is the boss. By that we mean that this temperature tells heat where and when to flow and, more generally, tracks the movement from non-equilibrium to equilibrium. In the previous section, when we described a counterexample to Clausius' version of the Second Law, we simply described a case where $T_G$ and  $T_L$ were out of sync. The violation occurred because heat was ``listening" to $T_G$ and not $T_L$. $T_G$ is the boss.

\begin{figure}[ht]
\centering
\includegraphics[scale = 0.0275, alt = {Diagram illustrating the Tolman effect. A rectangular ice block labeled $T_G$ = $T_L$, indicating thermal equilibrium at uniform potential, points via an arrow to a vertically oriented column with temperature layers $T_L3$, $T_L2$, $T_L1$. A downward arrow labeled $\phi$ represents gravitational potential. The image shows that while $T_G = T_L$ at the same potential, temperature varies with height when the system spans different potentials.}]{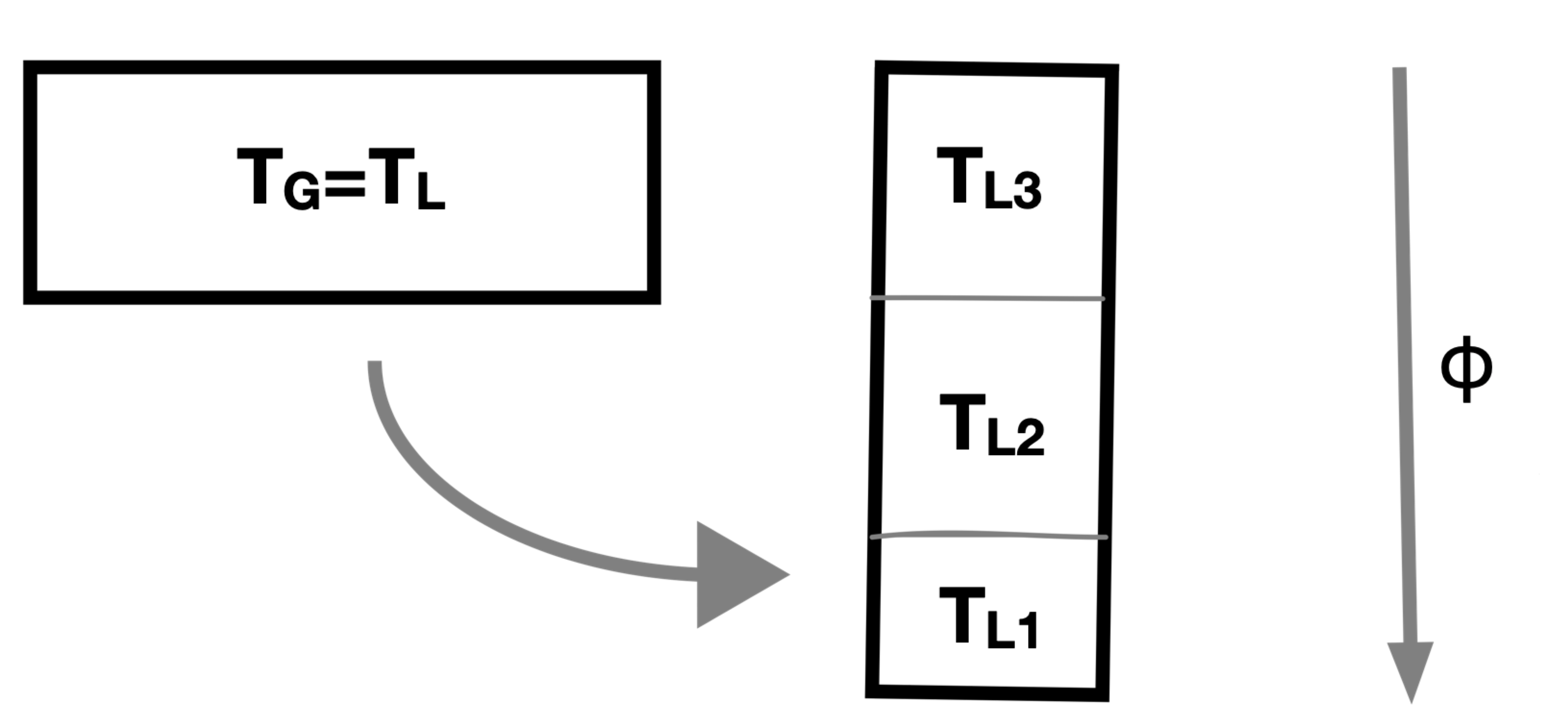}
\caption{Ice cube meets the Tolman effect.}
\label{fig:icecube}
\end{figure}

Here is another case to consider, inspired by a discussion in Balazs. Suppose we have a long block of ice laying horizontally at the same value of gravitational potential throughout. It is at the cusp of a phase transition to water, i.e., $0$ $^\circ C$. Now stand this block of ice up vertically, so that it experiences varying levels of gravitational potential. The parts of the block closer to the ground now have higher $T_L$ than before. (See Fig. \ref{fig:icecube}.) Suppose we rig things up such that $T_L$ is now high enough at the bottom to push past the melting point. Does the bottom melt? Well, if $T_G$ is still constant throughout we don't expect any melting. In terms of $T_L$ we would  have to change the melting point of water, making it depend explicitly on the local gravitational energy. 

The Tolman effect, we thus see, drives a sharp wedge between theory and observation as regards temperature. $T_G$ is perhaps the best choice of natural kind if one wants the simplest and most powerful generalizations of the facts of ordinary thermodynamics. After all, it saves the compact and powerful laws at the price of only a little complexity, and it varies with the core physical tendency underlying all this phenomena, the spontaneous movement of systems from non-equilibrium to equilibrium. That provides a powerful case for  $T_G$. However, in extended systems with varying gravitational potential, that is just not what you observe or measure. $T_L$ is. From this perspective $T_G$ `saved' thermodynamics only at the cost of making it empirically inaccessible to local observers. Thermodynamics was built upon measurements of the empirical temperature \parencite{chang_inventing_2004}. And the absolute temperature so crucial to the theory was vindicated by the empirical temperature successfully approximating it. True, the actual empirical temperatures always departed somewhat from the absolute temperature. But here we have a lawlike departure above and beyond the normal thermometer discrepancies: the more extended a system is, and the greater the gradient in gravitational potential, the further apart $T_G$ and $T_L$ get. That said, it's not clear that this discrepancy is worse than previous ones, e.g., that between air-thermometers and absolute temperature; after all, we know how $T_L$ approximates $T_G$.    

Either way, it seems like we must give up some parts of classical thermodynamics and the classical temperature: the physical meaning of temperature in terms of the transition from non-equilibrium to equilibrium or the local empirical accessibility of thermodynamics. 

\section{Clocks, Energies, Temperatures}

So far we've analyzed the temperature gradient of the Tolman effect in terms of energy and how it gravitates -- a standard view that even Einstein appeared to have, as we saw in \S2. We hear about ``the weight of heat" \parencite{tolman_temperature_1930, tolman_weight_1930}, ``the sedimentation of the heat energy in the gravitational field" \parencite{balazs_relativistic_1958}, the claim that ``heat has a weight, and heat has a tendency to sediment" that \parencite{balazs_thermodynamic_1965}, ``energy weighs" \parencite{haggard_death_2013}, or we are told that the energy of radiation has an ``equivalent mass density" which gravitates \parencite{santiago_tolman_2019}. These interpretations imply a genuine physical difference between systems at different $\phi$: a varying $T$ tracks intrinsic physical differences between parts of an extended gravitating system in equilibrium. Call this the \textit{energy interpretation} of the Tolman effect.

We see close ties between this energy interpretation of the Tolman effect and an energy interpretation of the gravitational redshift. That interpretation understands gravitational redshift in terms of ``the apparent weight of photons" \parencite{pound_apparent_1960} and how photons (e.g. a box of radiation) lose energy because they gravitate just like masses. As \textcite[p. 1]{okun_interpretation_2000} put it, sometimes ``this description is loosely phrased as a degradation of the photon's energy as it climbs out of [a] gravitational potential well". As seen in \S2, this readily leads to an energy gradient along varying $g_{00}$, resulting in a temperature gradient. Again, this seems to signal intrinsic physical differences between systems at different points of $\phi$.

It's instructive to see how the energy interpretation of redshift operates.\footnote{We follow \textcite{okun_interpretation_2000} here.} In Pound-Rebka-type experiments, photons are emitted upwards from a tower's base directly towards a receiver some height $h$ from the base. Empirically one observes them with a lower frequency (and hence energy) when measured at the top than at the bottom. A common way to explain this gravitational redshift of photons is to ascribe to photons a `relativistic mass' via their energy $E$ (photons are massless), $m_R = E/c^2$.\footnote{This follows from $E = m_R c^2$.} One uses this to say that massless particles gravitate because they also have `mass', contrary to classical physics. So, photons with energy $E$ -- as measured at Earth's surface -- are gravitationally attracted to Earth with force $gm_R$. Now we proceed with a classical analysis, that a photon's (kinetic) energy $E$ is converted into potential energy (via $\Delta \phi = -mgh$). If shot upwards away from Earth, where $\phi = 0$ on Earth's surface, photons lose some fraction of $E$ at height $h$ from Earth:
\begin{equation}
\frac{\Delta E}{E} = \frac{\Delta \omega}{\omega} = \frac{\Delta\phi}{m_Rc^2} = \frac{-m_Rgh}{m_Rc^2} = \frac{-gh}{c^2}.
\end{equation}
$\omega$ is the photon's frequency, and $g \approx 9.8m/s^2$. This is, up to a sign difference, the blueshift formula and so the energy interpretation appears to explain the Pound-Rebka results.

However, this leads to a puzzle. We know local clocks \textit{also} universally slow down, relative to a background `global' time, as they pass through $\phi$, or, equivalently, are uniformly accelerated. Local clocks -- the ticks of proper time $d\tau$ in local inertial frames\footnote{This is sometimes called the Clock Hypothesis, see e.g. Maudlin (2012, Ch. 5).} -- are related in static spacetimes to a global time coordinate $dt$ or `world clock' (definable because of the existence of the aforementioned global time-like Killing field) as $d\tau = \sqrt{g_{00}} dt$.\footnote{See \textcite{earman_gravitational_1980} for some subtleties.} As mentioned, we can approximate $g_{00}$ as $1 + 2\phi/c^2$ in the weak field limit (implicitly assumed in the energy interpretation when we use $\phi$). Thus, local clocks \textit{also} universally slow down (relative to unchanging $dt$) at higher $\phi$. But the observed frequencies of photons are measured in local inertial frames using local clocks counting proper time. Such clocks count the number of periods per unit proper time (i.e. the proper time period), and the observed frequencies are inversely related to the proper time period. Hence, the slowing ticks of proper time at different heights lead one to attribute a lower frequency, and hence decreasing energy, to the same system at different heights.

But now the puzzle arises. The Pound-Rebka experiments can be explained with \textit{either} of the two explanations above, but \textit{not both}: that results in (empirically inadequate) double-counting. Both predict the same observational outcome (the observed blueshift) but attribute it to different underlying mechanisms. So which is it: photons losing energy along increasing $\phi$, or clocks universally slowing down along increasing $\phi$? 

In a way, they are two sides of the same coin. But Okun et al argue that the energy interpretation is inappropriate for cleanly understanding redshift from the perspective of general relativity. A free-falling photon -- only under the influence of gravity -- is a perfect case of Einstein's equivalence principle: its behavior should be locally identical to that of an inertially moving photon, where nothing intrinsic to the photon has changed. 

This is supported by observing that the covariant derivative of the stress-energy tensor vanishes:
\begin{equation}
\nabla_\mu T^{\mu \nu} = 0 \iff \partial_\mu T^{\mu\nu} = - \Gamma^\mu_{\mu\sigma} T^{\sigma\nu} - \Gamma^\nu_{\mu\rho} T^{\mu\rho}.
\end{equation}
This can be written as:
\begin{equation}
\partial_\mu T^{\mu\nu} = - \Gamma^\mu_{\mu\sigma} T^{\sigma\nu} - \Gamma^\nu_{\mu\rho} T^{\mu\rho}
\end{equation}
where the Christoffel symbols, given by $\Gamma^\mu_{\nu \lambda} = \frac{1}{2} g^{\lambda \sigma} \left( \partial_\mu g_{\nu \sigma} + \partial_\nu g_{\mu \sigma} - \partial_\sigma g_{\mu \nu} \right)$, tracks how the metric $g_{\mu\nu}$ changes along a direction. This indicates that any apparent changes to a system's stress-energy is entirely due to changes in $g_{\mu\nu}$ encoded by $\Gamma^\lambda_{\mu \nu}$.

These support a \textit{clock interpretation} of the redshift -- local clocks (the rate of proper time) universally slow down as they go through increasing $g_{00}$, but the system's intrinsic properties remain unchanged. Interestingly, while \textcite[384-385]{einstein_doc_1911}'s discussion of a variable $c$ -- and the gravitational redshift -- began with an analysis steeped in terms of the energy interpretation as discussed in \S2, it actually ended in terms of \textit{clocks}: the clocks above and below "do not both give the "time" correctly." Since nothing happens to a free-falling photon (a stationary process), local observers see photons with a constant frequency, energy, and $c$ -- this is just the equivalence principle. 

However, when one attempts to compare their energy measurements with those made at a different $\phi$, i.e. a relatively accelerated system, one must recognize that $c$ is \textit{not} constant \textit{far away} from their local observations in the accelerated frame.\footnote{Here and in what follows, `far away' or `distant' should be understood in terms of distances relative to some scale determined by significant variation in the gravitational potential.} From this follows a ``consequence which is \textit{of fundamental significance for our theory} ... we must use clocks of unlike constitution for measuring time at places with differing gravitation potential" \parencite[106, emphasis ours]{einstein_doc_1911}. By ``unlike constitution", Einstein isn't suggesting that distant clocks must somehow be \textit{made} differently. Rather, he simply means that we must use clocks that tick at different rates (relative to clocks where we are, even if both are identically made) for measuring time at distant places with a different gravitational potential.\footnote{Einstein elaborates on the locution ``of unlike constitution" immediately after introducing it: ``For measuring time at a place which, relatively to the origin of the co-ordinates, has the gravitation potential $\Phi$, we must employ a clock which when removed to the origin of co-ordinates -- goes $(1 + \Phi)/c^2$) times more slowly than the clock used for measuring time at the origin of co-ordinates."} For two clocks to be of like constitution, then, is just for them to tick at the same rate.

Einstein's interpretation is conceptually continuous with the key insight of special relativity. In special relativity, $c$ is constant in all inertial frames, which lets us define time and space coordinates in terms of an unchanging `spacetime ruler' (giving us the Minkowski metric). In accelerated frames, $c$ remains constant \textit{at the origin} of any coordinate frame, i.e. to local observers. However, from the perspective of local observers who are \textit{accelerating}, or, equivalently, free-falling in the presence of $\phi$, $c$ is not necessarily constant away from the origin. If so, their `spacetime ruler' is \textit{not unchanging over spatial distances}; clocks vary over spatial distances. 

For instance, relative to the towertop with higher $\phi$, clocks at the bottom run faster by $\left[1 + \frac{\phi}{c^2} \right]$. Hence, photons emitted from below always appear to have higher frequency, and energy, to observers above than to observers below. But the photon itself has not undergone any changes -- it is undergoing a `stationary process'. This explains the Pound-Rebka effect without appealing to photons losing energy as a real physical change, moving us away from the energy interpretation, towards Einstein's interpretation that clocks universally slow down in the presence of a gravitational field. It also shows us that the redshift of radiation is just one \textit{specific instantiation} of a more general phenomenon: a variable `spacetime ruler' for accelerated observers looking far away. 

Common ways of presenting the Tolman effect also implicitly assume an energy interpretation when they talk about energy gravitating, or how systems in regions of lower gravitational potential are `hotter'. But this interpretation leads to a puzzle to do with the aforementioned violations of classical thermodynamics. The energy interpretation suggests that extended gravitating systems in equilibrium genuinely have an energy -- and hence temperature -- gradient that tracks something intrinsic about the system -- say, the amount of extractable work, when $T_L$ is higher at some point than another. After all, this is what the classical temperature tracked. Yet, as we've argued, this temperature gradient doesn't track any physically relevant features, such as the lack of heat flow between regions of varying temperature. 

On the clock interpretation, the Tolman effect simply follows from the general fact that \textit{there is no one set of universally applicable local time coordinates everywhere throughout the system in the presence of gravity}, contrary to special relativity. That is, from any local observer's perspective, distant clocks measure a different time -- one that is slowed down relative to local clocks. In this sense, they are of ``unlike constitution" in Einstein's terminology. Analogously, distant local thermometers measure $T_L$ differently given the Tolman effect. Hence, energy -- and temperature -- measurements from distant systems cannot be naively compared with measurements on local systems until these differences are corrected for.

The clock interpretation also draws attention to something we've been appealing to: the existence of a static global time coordinate $dt$ -- and the `world clock', with which comparisons of $d\tau$ at different spatial points is made possible. This clock is associated with \textit{global} time-translation symmetry of static spacetimes and the existence of the aforementioned time-like Killing field. Crucially, this clock is \textit{not} a local clock used by local observers. As mentioned, we can define global (kinetic plus gravitational) energy with respect to the global clock, along which it is conserved. Hence, a corresponding `world thermometer' would also measure a \textit{constant} global temperature, \eqref{tglobal}, throughout the system (by defining $T_G$ as the derivative of global energy with respect to entropy). However, this temperature is locally inaccessible because we never observe the absolute value of $g_{00}$; likewise, local observers never see the `ticks' of $dt$, only that of $d\tau$. 

\section{Truly Local Temperature?}

The clock interpretation leads to a \textit{third} interpretation of temperature, something that we've not seen proposed elsewhere regarding the Tolman effect. This is a radical proposal, but we see no reason to rule it out \textit{a priori}. Recall that there is no one set of universally applicable local time coordinates everywhere throughout the system in the presence of gravity. But since local energy measurements are made with local clocks, so are measurements of $T_L$ made by local `pocket thermometers'. If so, it seems that distant thermometers, like distant clocks, are also ``of unlike constitution", in that local thermometers simply measure something \textit{different} from distant ones in the presence of a varying gravitational potential. This leads us to the possibility that \textit{there is likewise no one set of universally applicable local temperature concepts in the presence of gravity}. On this proposal, temperature should be indexed to clocks and thermometers of `like constitution' but such indexed-temperatures need not be universally applicable. Given this interpretation, it's not so surprising that distant thermometers may be related to local thermometers in a way that reports local temperatures violating classical thermodynamics, as in the standard interpretation of the Tolman effect: all along they were tracking incompatible and distinct temperature concepts. Instead of trying to compare temperatures directly using thermometers of `unlike constitution', then, this third proposal suggests that we only use clocks and thermometers of `like constitution' with respect to which one defines local energy and local temperature.

We can make the proposal more precise by introducing the idea of \textit{local equilibrium frames}. The local equilibrium frame is, first and foremost, a local inertial frame: a frame defined relative to a point of an extended system in which the point of the system is momentarily \textit{at rest} with respect to its environment, and where spatial variations of $g_{00}$ are negligible in a sufficiently small neighborhood surrounding that point. However, it is a particular local inertial frame in which the system, at that point, \textit{also} satisfies certain thermodynamic conditions and is in \textit{thermodynamic equilibrium} with the environment, if only instantaneously; only then does it make sense to assign a temperature to this point of the system.\footnote{In relativistic hydrodynamics, there are two standard, if sometimes incompatible, ways of imposing such thermodynamic conditions. The \textit{Landau-Lifshitz energy frame} imposes the condition that the energy flux locally vanishes, while the \textit{Eckart particle frame} \parencite{eckart1940} imposes the condition that the particle flux is diffusion-free. See \textcite{akihiko2019} or \textcite{kovtun_temperature_2023} for discussion.} Since $g_{00} \approx 1$ in such frames, it is clear that the Tolman effect \eqref{tlocal} (approximately) vanishes for the region of that system which can be (approximately) described in terms of such frames, so that the entire region can be described by a constant local temperature.

Relative to such local equilibrium frames, we can now define a temperature: for each point of a system, define a `wahre-local' (truly local) temperature $T_{WL}$ indexed to that point's local equilibrium frame. Then, only make comparisons of $T_{WL}$ measurements for two systems to the extent that one can ignore variations in $\phi$ or $g_{00}$ for both systems, while still treating them in thermodynamic equilibrium. Simply put, define $T_{WL}$ indexed to local equilibrium frames. Since one can always choose a local inertial frame at a point and a small enough region around said point such that $g_{00} \approx 1$ to first approximation, and any system that can be ascribed a temperature should satisfy appropriate thermodynamic conditions, we can always define $T_{WL}$ relative to such a local equilibrium frame, for systems of interest. To the extent that we can ignore $g_{00}$ in the region around the point, to that extent we can employ a unique $T_{WL}$ concept in that region.

Physically, we can motivate $T_{WL}$ this way: this local equilibrium frame associated with a point picks out exactly the frame in which that point of the system is \textit{locally} at rest and is in thermodynamic \textit{equilibrium}, just as the global clock picks out the frame in which the \textit{entire extended system} is at rest and \textit{globally} in equilibrium. Then, to the extent that one can stomach $g_{00} \approx 1$, to that extent one can meaningfully talk about $T_{WL}$ and local thermodynamics. But $T_{WL}$ defined relative to different local equilibrium frames, at different points of an extended system, need \textit{not} be commensurate, unlike $T_{L}$; from this new perspective, they are just \textit{different} notions of temperatures, not to be na\"ively compared with each other.

To appreciate that this is indeed a third understanding of temperature, let's note the differences between $T_L$, $T_{WL}$, and $T_G$. Firstly, $T_L$ and $T_{WL}$ agree when $g_{00}$ can be assumed unchanging, but disagree about the temperature of distant systems. $T_L$ can be defined across varying $g_{00}$, at the cost of the breakdown of classical thermodynamics. To the contrary, $T_{WL}$ will be undefined insofar as $g_{00}$ isn't (approximately) constant. Secondly, $T_G$ and $T_{WL}$ can be defined for patches of spacetime, but differ in the sorts of patches required. $T_G$ can be applied insofar as we can find a patch of spacetime which can be approximated as a static spacetime, e.g. Schwarzschild spacetime, in which a global timelike Killing field exists. In contrast, $T_{WL}$ can be defined for any small enough region around any point of spacetime, for which local equilibrium frames are available. Furthermore, $T_{WL}$ is defined only relative to the local equilibrium frame at a point (insofar as it is valid), but this will in general not be identical to $T_G$ unless we are working with a flat spacetime (since there will be no Tolman effect otherwise).

If we employ $T_{WL}$, defined relative to local equilibrium frames, as the tracker of thermodynamic behavior, then it provides one way to live with the Tolman effect. Using local point-relative definitions of thermodynamic properties, $T_{WL}$ is indeed constant throughout the system whenever $T_{WL}$ is defined. Indeed we preserve all of classical thermodynamics relative to $T_{WL}$. Again, there are costs. Notably, we can't always define a universal $T_{WL}$ throughout an extended system -- when we cannot ignore second-order corrections to $g_{00}$, say. There is just no way to avoid relative differences in clock behavior across a varying gravitational potential. We simply accept that we've hit the limits of thermodynamics when we go beyond the local equilibrium frame. 

Interestingly, in a cognate debate about the \textit{special} relativistic temperature, Einstein had proposed something similar, dubbed the `eigen-temperature'. There, the question was whether there is a natural extension of the classical temperature concept to special relativity, cashed out in terms of whether there is a unique Lorentz transformation of the temperature (see e.g. \textcite{chua_t_2023}). Many proposals were made for whether a moving body appears hotter or cooler to a stationary observer, but Einstein suggested, late in the debate, that:
\begin{quote}
The actual problem is the treatment of the concept of temperature and heat [...] I am tempted to understand under the notion 'temperature' the reading from a comoving thermometer, i.e. to treat the temperature in every case as an invariant. (Einstein 1953, in \textcite[201]{liu_einstein_1992})
\end{quote}
Like our wahre-local temperature, Einstein's eigen-temperature is \textit{frame-relative}. It is defined \textit{relative to the co-moving / rest frame of the moving system}, i.e., the frame in which the system is in equilibrium.\footnote{This was later echoed by e.g. \textcite{landsberg_does_1966}.} It is also generally not measurable -- and hence incomparable -- over distances. The only way to measure the eigen-temperature is to \textit{go to the system's co-moving frame} -- the equilibrium frame. On this definition, the temperature of a distant object \textit{just is} its temperature in the co-moving frame, and hence is invariant by definition regardless of the frame from which you are looking at the system. Einstein's proposal thus eschews the possibility of measuring or observing the temperature from afar; a moving body's temperature just is its rest temperature, neither hotter nor cooler. 

Likewise, our proposal for $T_{WL}$ eschews the very same possibility. Given such a view, the variations of $T_L$ -- i.e. the Tolman effect -- should be interpreted similar to the redshift: it is a distortion of $T_{WL}$ because of our relative spatial distance to the point in question, caused by a varying gravitational potential, or, more generally, curved spacetime. We should not use it to attribute any intrinsic thermodynamic properties (or changes in them), any more than we should literally take redshifted energies/frequencies of incoming radiation to tell us the literal energies/frequencies at the source. From any one point, the temperature of a distant relatively accelerating point just is its rest temperature, defined in relation to \textit{the local equilibrium frame at that distant point}.

We are \textit{not} saying that $T_{WL}$ is the `true' or `best' successor to the classical temperature in general relativity. It's unclear to us whether there is a context-independent `true' or `best' natural counterpart in relativity to the classical temperature. However, we do think $T_{WL}$ is an interesting and \textit{valid} extension worthy of consideration. 

\section{Conclusion: The Robustness and Fragility of Thermodynamics}

\textcite[33]{einstein_autobiographical_1946} famously proclaimed that classical thermodynamics ``is the only physical theory of universal content concerning which I am convinced that, within the framework of the applicability of its basic concepts, it will never be overthrown.” But what are the limits of this framework's applicability? As he confronted the conceptual puzzles at the heart of special relativistic thermodynamics, he came to realize that
\begin{quote}
...there is actually no compelling method in the sense that one view would simply be 'correct' and another 'false'. One can only try to undertake the transition as naturally as possible. (Einstein 1953, in \textcite[200]{liu_einstein_1992})
\end{quote}
The issue here was that different methods for generalizing classical thermodynamics to special relativistic thermodynamics -- for instance, how to conceptualize the Carnot cycle and heat/work exchange in special relativity -- led to disagreeing Lorentz transformations for the temperature. Einstein's thought, then, was that we should find the most natural way to generalize the temperature, given that various methods appear ``correct" yet prescribe very different Lorentz transformations for the temperature.\footnote{See \textcite{chua_t_2023} for discussion.}

One proposed way to understand the natural extension of a concept to a new domain is via the \textit{consilience} of theoretical roles played by this concept \parencite{chua_t_2023}. To consider a framework's naturalness in some context, and whether it can be extended to a new context, we investigate whether the various theoretical roles played by its central concepts converge in some context, and whether (suitable extensions of) these theoretical roles continue to converge on a unique extension in the new context. 

In the local equilibrium frame of thermodynamic systems, the laws and quantities of thermodynamics are remarkably consilient; as \textcite{chua_t_2023} discusses for the classical temperature, the theoretical roles it plays -- its relation to heat in Carnot cycles, as an observable measured by thermometers, its relation to kinetic energy in kinetic theory, and its relation to frequencies in black-body radiation -- converge, capturing different aspects of the classical temperature's physical meaning, from the theoretical to the operational. This agreement picks out a robust regime in which reality can be described and manipulated, in which engines and air-conditioners are rendered possible, in which the laws of classical thermodynamics hold. However, what we've shown is that this regime -- the equilibrium regime -- is simultaneously fragile, depending on many contingent factors being `just right' for us to discover it. There is no guarantee that this regime is universally applicable, in the sense of finding a single `natural' successor in every domain. 

Chua has argued that this consilience of temperature breaks down under Lorentz boosts, and there is no natural successor to the temperature concept in relativity. Thermodynamics might not be a natural description for systems in relative motion to observers. But that's fine: in terrestrial thermodynamics, we (and our instruments) are approximately at rest with respect to the system of interest, and are always able to observe thermodynamic behavior. 

Likewise, we've argued here that thermodynamics appears to fragment in accelerated frames, or, alternatively, in a spatially varying gravitational field. The classical temperature can be generalized in at least three ways : $T_L$, $T_G$, or $T_{WL}$. $T_L$ preserves empirical accessibility to local observers, at the cost of many principles of thermodynamics. $T_G$ preserves these principles at the cost of empirical accessibility to local observers. Both assume universal applicability of the thermodynamic framework. $T_{WL}$ preserves empirical accessibility and the principles of thermodynamics by resisting its universal application beyond the local equilibrium frame. All three capture some desirable aspects of the thermodynamic framework, but not others; none recovers all that we might want from the classical temperature. We simply can't have it all. Thermodynamics might not be a natural description for relatively accelerated systems. 

What about distant systems, with which we are not at rest, for which we cannot ignore gravity? Even here we don't necessarily need relativistic thermodynamics. We can still define thermodynamic quantities \textit{in the equilibrium frame of the system}, correcting for the distance between us (e.g. via Doppler or gravitational redshift considerations). This might be an equilibrium frame with respect to local time, or an equilibrium frame with respect to global time (if we can idealize said system with a static spacetime). For example, to ascribe a temperature to a distant planet is to ascribe it a temperature \textit{as it is understood to be in equilibrium}, i.e. in \textit{its} equilibrium frame. In such cases we can generalize our application of thermodynamics, but \textit{only to a certain extent} because an equilibrium frame is still necessary. 

We thus echo Einstein's stance. There is, to us, `no compelling method' in the sense that one view of relativistic temperature is simply `correct' and another `false'. Each has its costs and benefits, and none wholly preserves the consilience of classical temperature. There is no natural relativistic temperature the same way there is a natural classical temperature. Nevertheless, that doesn't stop us from doing classical thermodynamics on relativistic systems as long as we can find equilibrium frames for them. We just need to be careful about which clock -- and temperature -- we deploy. 

In a way, this shouldn't be surprising. A basic lesson in general relativity is that `time' fragments in a world with variable curvature \parencite{callender_what_2017}. Each body has its own clock ticking off proper time; but, sometimes, locally or approximately, we can define additional natural time parameters like the global time. Since equilibrium thermodynamics requires a choice of time -- with respect to which systems stay in equilibrium -- and general relativity has many options, the fragmentation of time leads to a fragmentation of temperature.\footnote{\textcite{rovelli_thermal_2011} and \textcite{haggard_death_2013}'s ``thermal-time" proposal defines a universal time step at equilibrium, relative to Gibbs states $\rho$ representing systems in thermal equilibrium at temperature $T$. However, what we've discussed here suggests a fragmentation of the idea of equilibrium, and hence temperature: the system can be globally in equilibrium relative to the world clock $dt$, with corresponding global temperature $T_G$, or locally in equilibrium relative to proper time $d\tau$, with corresponding local temperatures $T_L$ or $T_{WL}$. This means that there is a corresponding fragmentation of the appropriate Gibbs state of interest: a global Gibbs state $\rho_G = e^{-\frac{1}{k\,T_G}H_G}$ with a global energy which includes gravitational degrees of freedom, or a local Gibbs state $\rho_L = e^{-\frac{1}{k\,T_L}H_L}$ with local energy which excludes gravitational degrees of freedom. This, in turn, means that there are different possible thermal times one can define. Of course, it's crucial to note that constructing a global Gibbs state -- and the statistical mechanics of gravity -- is an open problem; as Vidotto (2024) observes, extending statistical mechanics to spacetime itself is still work in progress.}


\numberwithin{equation}{section}
\numberwithin{figure}{section}
\numberwithin{table}{section}
\section*{Acknowledgements}

We thank Yichen Luo, Wayne Myrvold, Sai Ying Ng, John D. Norton, Carlo Rovelli, David Wallace, and audience members at the National University of Singapore Philosophy Department Speaker Series, the Southern California Philosophy of Physics Group Meeting, the Western Ontario Foundations of Thermodynamics Reading Group Meeting, and the UC Davis Philosophy of Physics Reading Group Meeting, for their comments and suggestions. We also thank two anonymous referees for their helpful feedback. 

\begin{flushright}
\emph{
Eugene Y. S. Chua \\
School of Humanities\\ 
Nanyang Technological University\\
Singapore, Singapore \\
eugene.chuays@ntu.edu.sg 
}
\end{flushright}
\begin{flushright}
\emph{
Craig Callender \\
Department of Philosophy \\
Institute for Practical Ethics\\
University of California, San Diego\\
La Jolla, CA, USA\\
ccallender@ucsd.edu\\
}
\end{flushright}

\printbibliography[title ={References}]

\end{document}